\documentclass[conference,usenames,dvipsnames]{IEEEtran} %
\IEEEoverridecommandlockouts
\usepackage{cite}
\usepackage{amsmath,amssymb,amsfonts}
\usepackage{algorithmic}
\usepackage{pgf}
\usepackage{tikz}
\usepackage{url}
\usepackage{hyperref}

\usetikzlibrary{shapes}
\usepackage[]{xcolor}
\usepackage{textcomp}
    
\hypersetup{
    pdftitle={RIGOLETTO - RIemannian GeOmetry LEarning : applicaTion To cOnnectivity\\A contribution to the Clinical BCI Challenge - WCCI2020},    % title
    pdfauthor={Marie-Constance Corsi, Florian Yger, Sylvain Chevallier, Camille Noûs},     % author
    pdfnewwindow=true,      % links in new window
    pdfkeywords={Riemannian geometry, functional connectivity, ensemble learning, BCI}, % list of keywords
    colorlinks=true,        % false: boxed links; true: colored links
    linkcolor=blue,       % color of internal links
    citecolor=blue,       % color of links to bibliography
    urlcolor=blue         % color of external links
}
\begin{document}

%\date{July 24., 2020}

\title{RIGOLETTO - RIemannian GeOmetry LEarning : applicaTion To cOnnectivity\\A contribution to the Clinical BCI Challenge - WCCI2020
%\thanks{}
}

% TODO 
% V2.0
% @MC - illustration du connectome pour un sujet (6 ?) -> t-test entre les deux conditions (pour les trois métriques)

\author{\IEEEauthorblockN{Marie-Constance Corsi}
\IEEEauthorblockA{\textit{Inria Paris, Aramis project-team} \\
\textit{Paris Brain Institute}\\
Paris, France \\
marieconstance.corsi@icm-institute.org}
\and
\IEEEauthorblockN{Florian Yger}
\IEEEauthorblockA{\textit{LAMSADE} \\
\textit{PSL, Univ. Paris-Dauphine}\\
Paris, France\\
florian.yger@dauphine.fr}
\and
\IEEEauthorblockN{Sylvain Chevallier}
\IEEEauthorblockA{\textit{LISV} \\
\textit{Univ. Paris-Saclay}\\
Versailles, France\\
sylvain.chevallier@uvsq.fr}
\and
\IEEEauthorblockN{Camille Noûs}
\IEEEauthorblockA{\textit{Cogitamus} \\
\textit{CNRS}\\
Paris, France\\
camille.nous@cogitamus.fr}
}

\maketitle
\date{July 24., 2020}

\begin{abstract}
This short technical report describes the approach submitted to the Clinical BCI Challenge-WCCI2020. This submission aims to classify motor imagery task from EEG signals and relies on Riemannian Geometry, with a twist. Instead of using the classical covariance matrices, we also rely on measures of functional connectivity.
Our approach ranked 1\textsuperscript{st} on the task 1 of the competition.
\end{abstract}

\begin{IEEEkeywords}
Riemannian geometry, functional connectivity, ensemble learning, BCI
\end{IEEEkeywords}

\section{Introduction}
Using a brain-computer interface (BCI) is a learned skill that requires time to reach high performance \cite{wolpaw_braincomputer_2002}. Despite its clinical applications \cite{pichiorri_brain-computer_2015, king_operation_2013}, one of the main drawbacks is the high inter-subject variability that could be noticed for performance. This is sometimes referred in the literature as the "BCI inefficiency" phenomenon \cite{allison_could_2010,thompson_critiquing_2018} and affects its usability. Among the approaches adopted to tackle this issue are the search for neuromarkers, that potentially capture better the neurophysiological mechanisms underlying the BCI performance \cite{blankertz_neurophysiological_2010-1,ahn_high_2013}, and the optimization of classification pipelines \cite{lotte_review_2018}, that could be robust enough to be applied to any subject. 

In this work, we proposed an original approach that combines functional connectivity estimators, Riemannian geometry and ensemble learning to ensure a robust classification. This article not only describes the proposed approach but also presents the methodology and the results that were conducted for our submission.

\section{Riemannian Geometry}
As pointed out in~\cite{lotte_review_2018}, the use of Riemannian geometry for Motor Imagery BCI is one of the breakthroughs of the last $10$ years of research in BCI and is now the golden standard.

\begin{figure}[t]
    \centering
    \includegraphics[scale=0.5]{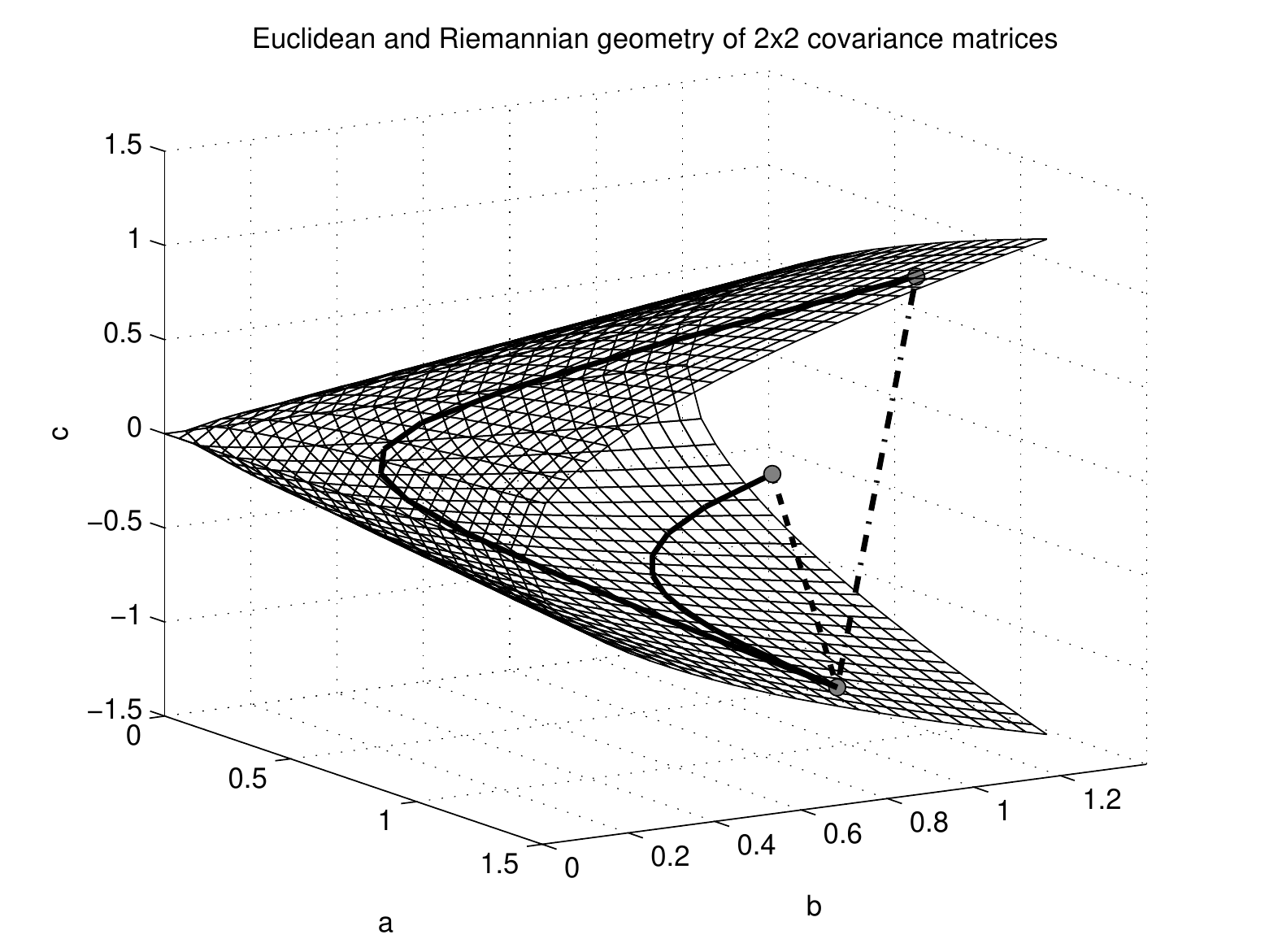}
    \caption{Comparison of Euclidean and Riemannian geometries for $2 \times 2$ SPD matrices. }
    \label{fig:RG_PSD}
\end{figure}

\begin{figure*}[tb]
    \begin{center}
	\includegraphics[width=0.8\linewidth]{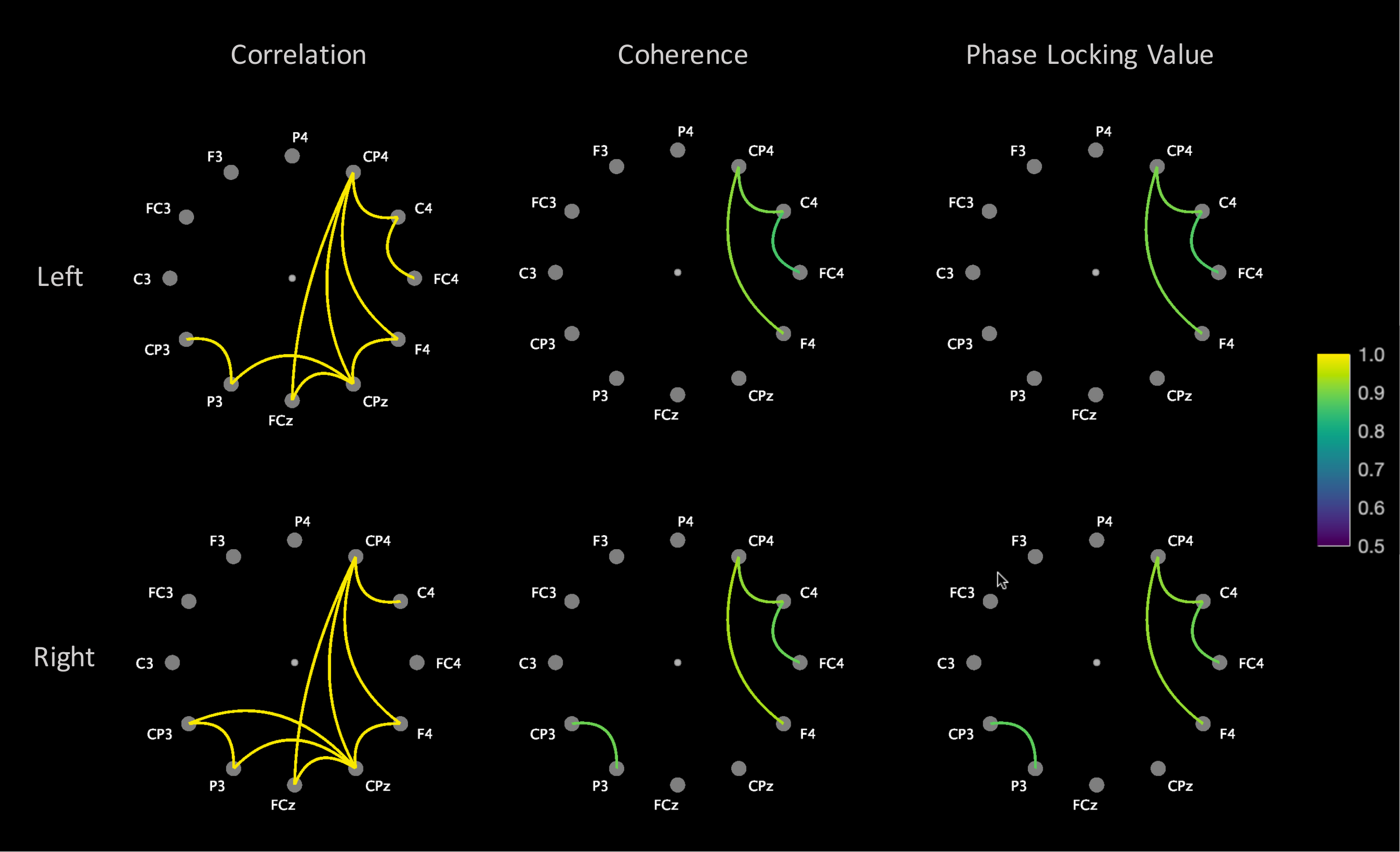}
	\caption{Features extraction. Each line corresponds to a given condition (left or right) and each column is associated with a given estimator: Correlation, Coherence and Phase Locking Value. We obtained these figures from subject 6 by averaging the estimators over the trials of the training set. In each case, the threshold $Th$ was obtained as follows: $Th=Min+0.9*(Max-Min)$, where Min and Max refer respectively to the minimum and the maximum values obtained for a given case (i.e. metric and experimental condition).}
	\label{Figure_FC}%
    \end{center}
\end{figure*}

The approach consists in extracting Symmetric Positive Definite (SPD) matrices that are symmetric matrices with strictly positive eigenvalues, usually the covariance matrices among sensors, for each epoch and then in considering this space as a curved (i.e. Riemannian) space.
As illustrated in Fig.~\ref{fig:RG_PSD} for $2 \times 2$ matrices, the space of SPD matrices could be considered as a Euclidean space (as a subspace of the Symmetric matrices) but several drawbacks occur (e.g. \textit{swelling effect} - see~\cite{yger2016riemannian}). Those drawbacks are leveraged when the Riemannian geometry is used and the distance between two SPD matrices $A$ and $B$ is expressed as :
\begin{align}
 \delta_R (A,B) = || \log \left( A^{-\frac{1}{2}} B A^{-\frac{1}{2}} \right) ||_\mathcal{F}
 \label{eq:rg_def}
\end{align}
with $\log (\cdot)$ the matrix logarithm and $||\cdot ||_\mathcal{F}$ the Frobenius norm.

However, in practice, we will favor another Riemannian geometry with similar properties but being faster to compute\footnote{The relationship between those geometries is developed in~\cite{yger2015supervised}.}, the LogEuclidean distance :
\begin{align}
 \delta_{LE} (A,B) = || \log \left(A\right) - \log \left(B\right) ||_\mathcal{F}
 \label{eq:LE_def}
\end{align}

In both geometries, the K\"archer average of a set of matrices $\{X_1,\cdots X_n \}$ is defined as : 
\begin{align}
\min_X \sum_{i=1}^n \delta^2 (X_i, X)
 \label{eq:karcher_avg}
\end{align}
A closed-form solution exists for $\delta_{LE}$ (but not for $\delta_R$)
\begin{align}
    \bar{X}_{LE} = \exp \left( \frac{1}{n} \sum_{i=1}^n \log\left(X_i \right)\right)
\end{align}

A simple, yet efficient classifier for SPD matrices consists in computing the K\"archer average of each class and then in predicting for a given test sample the class which average is the closest (using $\delta$).

The interested reader can refer to~\cite{yger2016riemannian,congedo2017riemannian} for more details. Until now, the Riemannian geometry was applied on SPD matrices extracted from covariances among sensors but other characteristics extracted from the EEG signal could produce SPD matrices. The next section will describe an alternative way to obtain SPD matrices based on functional connectivity.

\section{Functional connectivity}
Functional connectivity (FC), which consists of assessing the interaction between different brain areas ~\cite{de2014graph,bastos_tutorial_2016}, can be a valuable tool to provide alternative features to discriminate subjects' mental states \cite{cattai_phaseamplitude_2019} and to study neural mechanisms underlying BCI learning \cite{corsi_functional_2020}. 

Here, as an exploratory study, we considered complementary undirected FC estimators to assess which of them, associated to Riemannian geometry, could best classify the data. 
For a given FC estimator, we took into account a time window of [3, 7.5 s] and we averaged the FC values within the alpha-beta band [8, 30 Hz]. Computations were made using the Brainstorm toolbox~\cite{tadel_brainstorm:_2011}.
In the following subsections, we defined the metrics computed between two given signals referred as $s_{1}(t)$ and $s_{2}(t)$ between two EEG sensors. An illustrative example is presented in Fig. \ref{Figure_FC}.

\subsection{Spectral estimation}
We computed two spectral estimators: the coherence (Coh) and the Imaginary coherence (ICoh). Coh and ICoh are both computed from the coherency, defined as the normalized cross-spectral density obtained from two given signals. More specifically, they are obtained as follows:

\begin{equation}
Coh_{12}(f)=\frac{|S_{12}(f)|^2}{S_{11}(f).S_{22}(f)}\
\end{equation}

\begin{equation}
ICoh_{12}(f)=\frac{\Im{S_{12}(f)}}{\sqrt{S_{11}(f).S_{22}(f)}}\
\end{equation}
with $S_{12}(f)$ the cross-spectral density and $S_{11}(f)$ the auto-spectral density.

The advantage of ICoh is its reduced sensitivity to signal leakage and volume conduction effects \cite{nolte_identifying_2004,colclough_how_2016}.

\subsection{Phase estimation}
As a phase estimator method, we worked with the Phase Locking Value (PLV), which assesses phase synchrony between
two signals in a specific frequency band~\cite{lachaux_measuring_1999,tass_detection_1998, aydore_note_2013}. More specifically, it corresponds to the absolute value of the mean phase between $s_{1}$ and $s_{2}$, defined as follows:
\begin{equation}
PLV=|e^{i\Delta\phi(t)}|
\end{equation}
where $\Delta\phi(t)=arg(\frac{z_1(t).z_2^{*}(t)}{|z_1(t)|.|z_2(t)|})$\
\\
$\Delta\phi(t)$ represents the associated relative phase computed between signals and $z(t)=s(t)+i.h(s(t))$ the analytic signal obtained by applying the Hilbert transform on the signal $s(t)$.

\subsection{Amplitude coupling method}
We computed the Amplitude Envelope Correlation (AEC)
~\cite{ hipp_large-scale_2012, brookes_measuring_2011,colclough_how_2016} which relies on the linear correlations of the envelopes of the band-pass filtered signals obtained from Hilbert transform. 

For the sake of completeness, we report the results of both AEC and ICoh, although those features were not used in the final submission. The generated matrices were not SPD and we had to pre-process them heavily in order to be able to apply the Riemmannian geometry. This may explain the poor results of those features in our setup. 

\section{Proposed approach : RIGOLETTO}

The novelty of our approach consists of combining Riemannian classifiers trained on SPD matrices coming from both measures of FC and covariance estimation.

\subsection{Task 1 : within-subject classification}

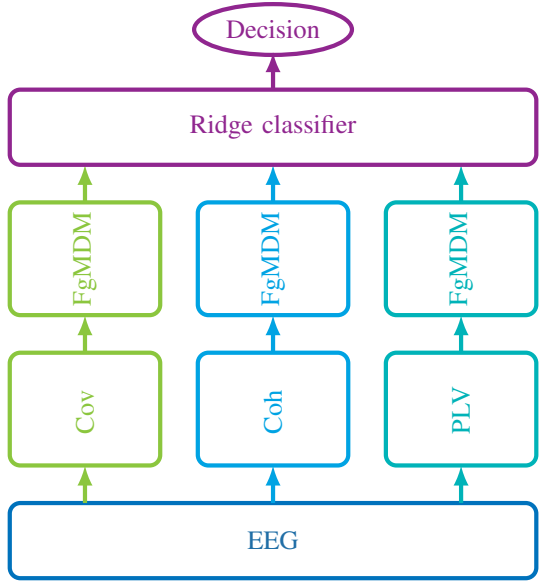
\begin{figure}[tb]
    \centering
    \begin{tikzpicture}
    \draw[ultra thick, color=RoyalBlue, rounded corners] (0, 0) rectangle (7, 1) ;
    \draw[color=RoyalBlue!80!black] (3.5, 0.5) node {EEG} ;

    \draw[>=latex, ->, ultra thick, color=LimeGreen] (1, 1) --  (1, 1.5) ;
    \draw[>=latex, ->, ultra thick, color=Cerulean] (3.5, 1) --  (3.5, 1.5) ;
    \draw[>=latex, ->, ultra thick, color=BlueGreen] (6, 1) --  (6, 1.5) ;
    
    \draw[ultra thick, color=LimeGreen, rounded corners] (0, 1.5) rectangle (2, 3) ;
    \draw[ultra thick, color=Cerulean, rounded corners] (2.5, 1.5) rectangle (4.5, 3) ;
    \draw[ultra thick, color=BlueGreen, rounded corners] (5, 1.5) rectangle (7, 3) ;
    
    \draw[color=LimeGreen] (1, 1.8) node[rotate=90,right] {Cov} ;
    \draw[color=Cerulean] (3.5, 1.8) node[rotate=90,right] {Coh} ;
    \draw[color=BlueGreen] (6, 1.8) node[rotate=90,right] {PLV} ;
    
    \draw[>=latex, ->, ultra thick, color=LimeGreen] (1, 3) --  (1, 3.5) ;
    \draw[>=latex, ->, ultra thick, color=Cerulean] (3.5, 3) --  (3.5, 3.5) ;
    \draw[>=latex, ->, ultra thick, color=BlueGreen] (6, 3) --  (6, 3.5) ;
    
    \draw[ultra thick, color=LimeGreen, rounded corners] (0, 3.5) rectangle (2, 5) ;
    \draw[ultra thick, color=Cerulean, rounded corners] (2.5, 3.5) rectangle (4.5, 5) ;
    \draw[ultra thick, color=BlueGreen, rounded corners] (5, 3.5) rectangle (7, 5) ;

    \draw[color=LimeGreen] (1, 3.5) node[rotate=90,right] {FgMDM} ;
    \draw[color=Cerulean] (3.5, 3.5) node[rotate=90,right] {FgMDM} ;
    \draw[color=BlueGreen] (6, 3.5) node[rotate=90,right] {FgMDM} ;
    
    \draw[>=latex, ->, ultra thick, color=LimeGreen] (1, 5) --  (1, 5.5) ;
    \draw[>=latex, ->, ultra thick, color=Cerulean] (3.5, 5) --  (3.5, 5.5) ;
    \draw[>=latex, ->, ultra thick, color=BlueGreen] (6, 5) --  (6, 5.5) ;
    
    \draw[ultra thick, color=Plum, rounded corners] (0, 5.5) rectangle (7, 6.5) ;
    \draw[color=Plum] (3.5, 6) node {Ridge classifier} ;
    
    \draw[>=latex, ->, ultra thick, color=Plum] (3.5, 6.5) --  (3.5, 7) ;
    
    \node[ultra thick, draw, ellipse, color=Plum] (s)at(3.5, 7.3) {Decision};
    %\draw[ultra thick, color=Magenta] (0, 5) --  (1, 5.5) ;
    %\draw[ultra thick, color=Magenta] (3.5, 5) --  (3.5, 5.5) ;
    %\draw[ultra thick, color=Magenta] (6, 5) --  (6, 5.5) ;

    \end{tikzpicture}
    \caption{Classification pipeline: coherence, phase locking value and spatial covariances are estimated from the EEG signal. A first level of classification was performed by FgMDM classifiers, that yielded output decision probabilities to train a second level classifier, a ridge regression classifier, that provided the final decision.}
    \label{fig:pipelines}
\end{figure}
 
To estimate FC features, we used the computation detailed in the previous section implemented in Brainstorm software~\cite{tadel_brainstorm:_2011}, and the sample covariance estimator of Matlab.
The covariance estimators often include regularization using shrinkage approach to avoid ill-conditioned matrices. For FC, no shrinkage estimators had been defined yet. Thus, we used a simple algorithm to project FC matrices on the manifold of PSD matrices~\cite{higham1988computing}.

\begin{figure*}[tb]
    \centering
    \pgfimage[interpolate=true,width=0.9\linewidth]{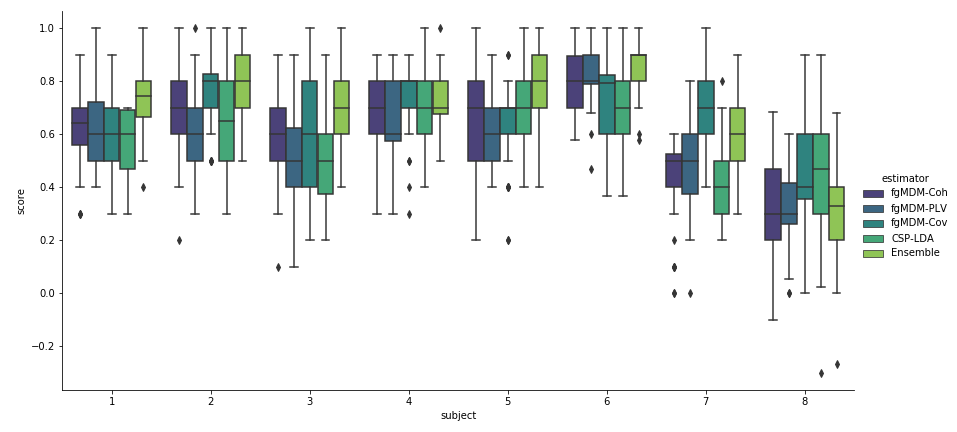}
    \caption{Per subject Kappa score, comparing the separate pipelines, i.e. FgMDM estimated on covariance, spectral coherence and PLV, a CSP+LDA for the baseline and the submitted ensemble classifier.}
    \label{fig:ens-subj}
\end{figure*}

\begin{figure}[tb]
    \centering
    \pgfimage[interpolate=true,width=\linewidth]{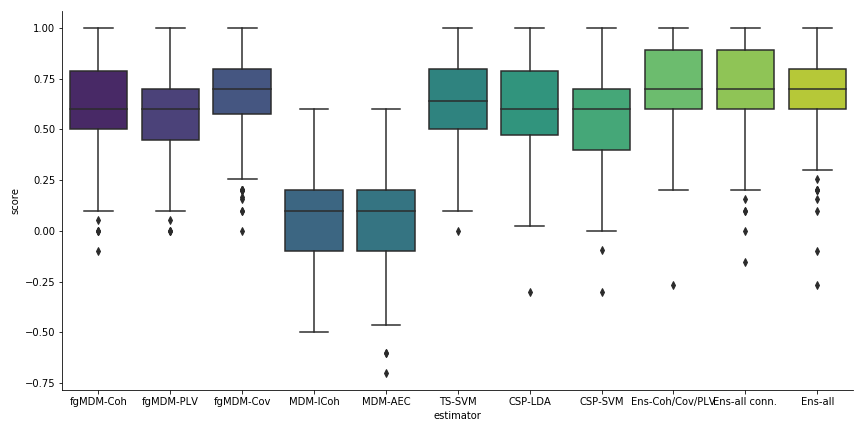}
    \caption{Average kappa score for all subjects, for the different tested estimators.}
    \label{fig:ens-pipelines}
\end{figure}

To classify FC and covariance matrices, we rely on Python~\cite{van1995python,oliphant_python_2007} and its libraries: Numpy~\cite{oliphant_guide_2006,walt_numpy_2011}, Scipy~\cite{virtanen_scipy_2020}, Pandas~\cite{mckinney_python_2012}, Scikit-Learn~\cite{scikit-learn}, MNE~\cite{gramfort_meg_2013},  Jupyter~\cite{kluyver_jupyter_2016}, Matplotlib~\cite{hunter_matplotlib_2007}, Seaborn~\cite{waskom_mwaskomseaborn_2018}. 
We applied the FgMDM algorithm~\cite{barachant2010riemannian}, that compute filters from a Fisher Geodesic Discriminant Analysis before using a Minimum Distance to Mean (MDM) classifier. We used the LogEuclidean distance and its associated mean in the MDM for its robustness and its efficiency. To take into account the shift between training and test set, the test data were transported on to the mean of training set\footnote{This transductive setup was allowed by the rules of the competition but in a real-life scenario, the mean of the test data could be estimated with unlabelled data during the calibration.}, as described in~\cite{barachant2011multiclass,yger2015supervised}.

Each FgMDM classifier predicted a probability for the output classes using the softmax function on distance to nearest mean. These probabilities were used to train a stacked classifier~\cite{wolpert1992stacked}. We tried several classifiers and we chose the ridge classifier for its robustness. This classifier made the final decision for the prediction, as shown in Fig.~\ref{fig:pipelines}.

The performance of our submission was estimated on training data and compared to a baseline, that was the Linear Discriminant Analysis LDA) with CSP spatial filters. The results are shown in Fig.~\ref{fig:ens-subj}, indicating the Kappa score estimated with repeated 5-fold cross-validation for each subject. The performance of each level 1 classifiers -- the FgMDM-Coh, FgMDM-PLV, FgMDM-Cov -- are provided, along with the ensemble classifier.

Indeed, we tested several classifiers and a combination for the stacked classifier. The obtained results are summarized on Fig.~\ref{fig:ens-pipelines}. 
The FgMDM trained on ICoh and AEC features presented very low kappa score. We also used a popular RG classifier, an SVM trained on the tangent space (TS-SVM). We also tested the CSP-SVM. In both cases, the SVM was parametrized through a grid search on the parameter space. 
This figure displays the score of the chosen system and stacked classifiers trained on different features: one ensemble classifier trained on all FC features (Coh, ICoh, PLV, AEC, Cov) and one trained on all possible level one classifiers. %(FgMDM trained on Coh, ICoh, PLV, AEC, Cov, TS-SVM, CSP-SVM and CSP-LDA).
For more details on the Task 1, the reader can refer to \cite{corsi2021rgfc}.

\subsection{Task 2 : across subjects decoding}

We computed the  K\"archer average for the data of each subject, that is on the training data for subjects $1$ to $8$ and on test data for subject $9$ and $10$. 
We chose then the ensemble classifier described above from the subject with the closest mean in the sense of the $\delta_R$ to predict the unknown labels of subjects $9$ and $10$ respectively.
We also built a voting classifier (data not shown), which combined the output of all the classifiers trained on each subject weighted with the inverse of the distance between K\"archer average. The results were less stable for subjects 1 to 8, thus we favored the simpler but more effective scheme of selecting only the classifier trained on the subject with the closer K\"archer average.

\begin{figure*}[tb]
    \centering
    \pgfimage[interpolate=true,width=\linewidth]{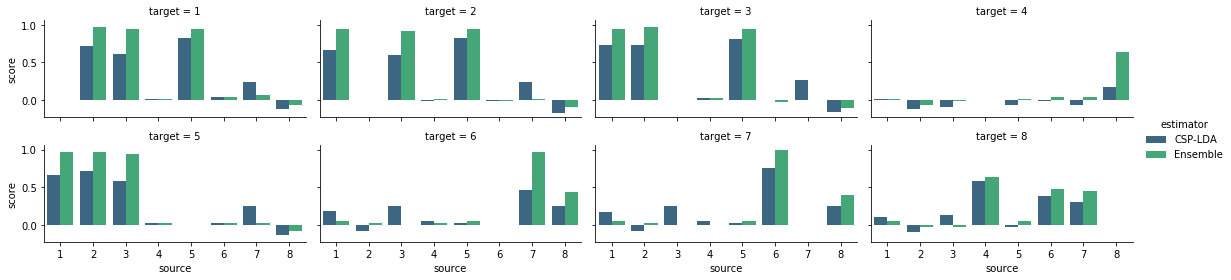}
    \pgfimage[interpolate=true,width=\linewidth]{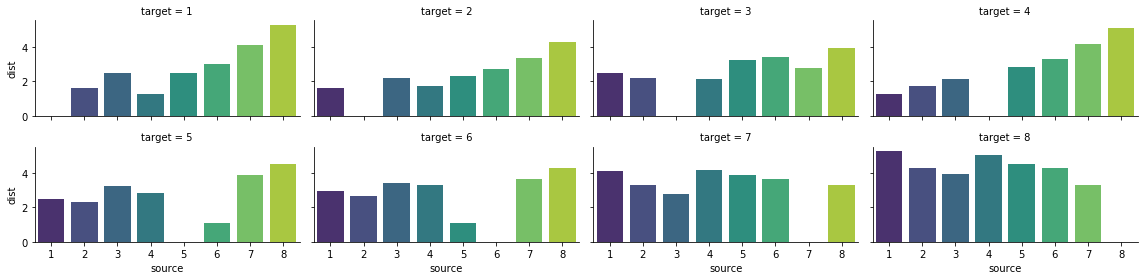}
    \caption{Top: Kappa score for each target subjects, comparing our system and CSP-LDA trained on each source subject. Bottom: Distance $\delta_R$ from average covariance matrix of source subject  $\bar{X}_{\text{src}}$ to average matrix of target subject $\bar{X}_{\text{tgt}}$.}
    \label{fig:ens-transfer}
\end{figure*}

To validate our system, we estimated the score with a leave-one-out scheme, that is training on all but one subject and computing score on the left-out subject.
The Fig.~\ref{fig:ens-transfer} shows the Kappa score estimated when trained on a given source subject and making prediction for a given target subject.

\begin{figure}[tb]
    \centering
    \pgfimage[interpolate=true,width=1\linewidth]{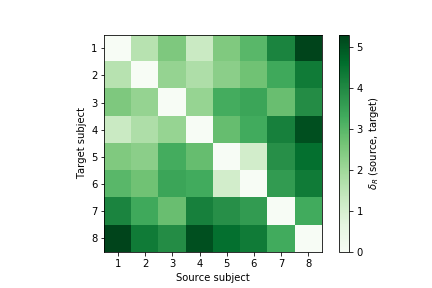}
    \caption{Distance between the K\"archer average of each target subject with each source subject.}
    \label{fig:dist-transfer}
\end{figure}

\subsection{Aftermath}
There were 14 submissions to the competition from 12 different institutions around the world across 9 different countries spread across 3 continents. 
At the end of the competition\footnote{For more details, the reader can access to the website of the competition~: \url{https://sites.google.com/view/bci-comp-wcci/}}, our submission ranked as described in Table~\ref{tab:finalscore}.

\begin{table}[t]
\begin{tabular}{|l|||l||l|l|l|}
\hline
                        & Kappa CV & Accuracy & Kappa & Rank \\\hline\hline
Within-subject (task 1) & 0.68     & 78.44\%  & 0.57  & 1    \\\hline
Cross-subject (task 2)  & 0.55     & 25.00\%  & -0.50 & 13   \\\hline
Overall                 & -        & -        & -     & 4   \\\hline
\end{tabular}
\label{tab:finalscore}
\end{table}

We mainly focused on task 1 and our approach got the first position on this task with a substantial margin, the following teams having respectively kappa scores of $0.49$ and $0.47$ and accuracies of $74.69\%$ and $73.75\%$. 
The kappa score obtained on validation is close to 0.68, the value obtained on training data with a 5-fold cross-validation, indicated in the Kappa CV column.

On the second task, we submitted a very simple approach and it seemed to have performed worse than random. This could mean that our approached matched patients having opposite patterns, hence leading to this result. 
Despite this inversion of class prediction, the absolute value of kappa obtained on the validation set is close to the one obtained on the training data (and the inverse classifier would have reached $75\%$ accuracy). This very curious phenomenon will be investigated in the near future.
The proposed approach could be refined by matching the means of each class (for known subjects) to a k-means (with $k=2$ and using K\"archer means) or adding weights for each subject in the spirit of~\cite{kalunga2018transfer}.

Rank aggregation\footnote{Note that in our case, the voters would be the tasks and the candidates would be the submissions.} has been the focus of many researchers in the field of computational social choice. The competition organizers choose to make the overall ranking using a weighted sum of the kappa value of the two tasks (the weight for a given competitor on a task being $(15-r)$ with its rank $r$). However, this rank aggregation technique is quite peculiar and unknown in the literature of computational social choice (where Kemeny optimal aggregation would be the classical way to go). First, the ranking is not really performed according to a weighted sum of the performances (as the weights are specific to each user) and it uses the ranking as weights (then, comparing two candidates could depend on the performance of a third one). Moreover, the rank aggregation method has been selected by the organizers \emph{a posteriori}\footnote{Note that we do not dispute our overall rank -as our approach performed quite differently on each task- but we would rather take this opportunity to discuss the ranking process for the sake of good order, this discussion being meant as a feedback competitions' organizers.} and knowing it, perhaps some competitors would have changed the focus of their submission.

For a full description of our pipeline and the requirements to use it, the reader can refer to the RIGOLETTO GitHub repository on \url{https://github.com/sylvchev/wcci-rgcon}. 

As previously explained, our main objective was to optimize the classification accuracy in the within-subject category. Using FC estimators associated with an ensemble classifier gives the possibility to take into account the users' specificity. 

After participating to the competition, we elicited different approaches to improve our method depending on the adopted perspective: theory behind RG, features extraction and transfer learning.
In the first case, further investigation should be done regarding the follow-up to the MDM~\cite{congedo2019riemannian} and the dimensionality reduction~\cite{horev2016geometry} (for other higher dimensionality datasets).
Regarding the feature extraction, we plan to improve in particular the selection of the frequency band of interest~\cite{klimesch_eeg_1999}. Another promising lead would be to extract for each epoch several PSD matrices, each on a different frequency band, and to consider this set as a trajectory on the manifold, in the spirit of~\cite{li2012electroencephalogram}. Other items, such as the agreement and variability among covariance and connectivity and the non-stationarity of connectivity features~\cite{balzi2015importance} will be considered.
Finally, we plan to study the impact of the centering operation on transfer learning tasks~\cite{rodrigues2018riemannian}.

Participating to the WCCI-Clinical BCI Competition has been the occasion to propose a novel approach and to start bridging the gap between Riemannian geometry and connectivity features. This resulted in a very practical algorithm and it brought promising results as well as intriguing failures. Nevertheless, it motivates the need to study more in depth the connectivity features under the lens of the Riemannian geometry. 
For instance, Fig.~\ref{Figure_FC} showed that covariance and connectivity features seems to produce similar average patterns but this raises as well the question of their individual variability. We plan to study those questions in the next future.

\section{Authors contributions}
Camille Noûs\footnote{For more information, see~\url{https://www.cogitamus.fr/camilleen.html}} is a collective individual and contributed to the collegial construction of the standards of science, by developing the methodological framework, the state-of-the-art, and by ensuring post-publication follow-up. This co-authorship symbolizes the collaborative nature of this work.
All the authors contributed equally to this work, taking advantage of their complementary skills. MCC was in charge of the data exploration and of the features extraction from functional connectivity. FY and SC were in charge of applying Riemannian geometry from FC estimators and designing the classification pipeline. All the authors wrote, revised and approved the submission.
Finally, MCC was chosen as a team leader and managed the submission of our approach and the communication with the organizers.

\section{Environmental impact}
The approach taken in this submission does not require lengthy computation on GPU clusters or HPC, in order to reduce its environmental impact. This submission generated the equivalent of 62 gCO$_2$, that is comparable to watching the whole \emph{``Lord of the Ring''} trilogy on an HD streaming service.

Training and experimenting with the models involved the equivalent of 6h of full load CPU computation. They were executed on a desktop computer with a 600W PSU that consumes 0.4 kWh during computation, measured with a wattmeter, and operated in France where the carbon footprint is 4.56 gCO2/kWh~\cite{ademe_bilan_2018}. The whole computation  generated the equivalent of 10.94 gCO$_2$.

The team members relied mainly on Slack, git and overleaf to communicate. As there is no direct estimation of the footprint of these services, we use the email scenario of The Shift Project report~\cite{ferreboeuf_lean_2019} as a surrogate. The digital action of sending an email is characterized by the 5 minutes use of a terminal plus 1 MB of transmitted data; it generates 0.3 gCO$_2$ according to The Shift Project. We evaluate that our submission required the equivalent of 170 mails following this scenario. The estimated footprint is thus 51 gCO$_2$.

This submission generated the equivalent of 62 gCO$_2$. The Shift Project made a contested estimation for the environmental impact of watching a video in HD on a streaming service~\cite{geist_did_2020}. While this is still debated,the cost is estimated to be circa 1 gCO$_2$ for 10 minutes of HD video. Our submission is thus somewhere between watching the theater-released version or the extended version of the Lord of the Ring trilogy on streaming.

% 6h * 0.4 kWh * 4.56 gCO2 (PC-manip - Base Carbone https://www.bilans-ges.ademe.fr) + (100msg slack + 50 git + 20 overleaf) * 0.3 gCO2 (Shift project Rapport_Pour-une-sobriété-numérique)

\section{Acknowledgements}
The authors would like to thank the organizers of the WCCI-Clinical BCI competition for giving them the opportunity to kickstart this promising collaboration. %to setup classification pipelines dedicated to BCI end-users.
FY would like to thank Fabien Lotte who suggested several years ago to investigate the use of connectivity features in lieu of the usual covariance estimator in Riemannian classifiers.

FY acknowledges the support of the ANR as part of the "Investissements d'avenir" program, reference ANR-19-P3IA-0001 (PRAIRIE 3IA Institute).
SC acknowledges that this work could have supported by ANR or IDEX but is only supported by the recurrent funding of the UVSQ.
%Also, this project could have been supported by the project R2B2 (Reliable Riemannian Brain-computer interfaces outside the laB) if it had been accepted and funded by the ANR in 2017...

\bibliographystyle{./bibliography/IEEEtran}
\bibliography{biblio}

\end{document}